\documentclass[letterpaper]{JHEP}

\usepackage{amssymb,amsmath}
\usepackage{epsfig}

\def\be{\begin{eqnarray}}
\def\ee{\end{eqnarray}}

\def\3{\ss}

\def\half {\frac{1}{2}}

\preprint{{\tt hep-th/0510076}}

\title{Semi-localized instability of the Kaluza-Klein linear dilaton vacuum}

\author{Oren Bergman and Shinji Hirano\footnote{Address starting October 10, 2006:
Niels Bohr Institute, Copenhagen University, Blegdamsvej 17, DK-2100 Copenhagen,
Denmark.}\\
Department of Physics\\
Technion, Israel Institute of Technology\\
Haifa 32000, Israel\\
\email{bergman, hirano@physics.technion.ac.il}}

\abstract{The Kaluza-Klein linear dilaton background of the bosonic string
and the Scherk-Schwarz linear dilaton background of the superstring are shown 
to be unstable to the decay of half of spacetime. 
The decay proceeds via a condensation of a semi-localized tachyon
when the circle is smaller than a critical size,
and via a semiclassical instanton process when the circle is larger
than the critical size.
At criticality the two pictures are related by a duality of
the corresponding two-dimensional conformal field theories.
This provides a concrete realization of the connection between 
tachyonic and semiclassical instabilities in closed string theory,
and lends strong support to the idea that non-localized closed string
tachyon condensation leads to the annihilation of spacetime.}


\begin{document}


\section{Introduction}

In a now classic paper, Witten demonstrated that the Kaluza-Klein (KK) vacuum 
of five-dimensional gravity is unstable to decay semiclassically via
a ``bubble of nothing'' \cite{Witten:1981gj}. 
The bubble is nucleated by an instanton process described
by the five-dimensional Euclidean Schwarzschild black hole, and 
then expands at a speed approaching the speed of light.
Witten's result has been generalized to other cases, such as the KK
magnetic field background \cite{Dowker:1995gb}, supergravity fluxbranes
\cite{Costa:2000nw}, twisted circle compactifications in supergravity 
\cite{DeAlwis:2002kp},
and in particular Scherk-Schwarz (SS) compactification.
The semi-classical description of the decay in terms of a gravitational
instanton is valid as long as $R\gg l_P$.
As the radius of the circle decreases the (super)gravity description
eventually breaks down, and must be replaced with string theory.
However one does not in general know the exact string backgrounds
corresponding to these instantons.

String theory in non-supersymmetric backgrounds often exhibits another
type of instability as well, namely one associated with tachyonic modes.
The stable background in these cases is given by a tachyon condensate.
Since this corresponds to an off-shell state in the original background,
we need to use off-shell techniques like string field theory and
worldsheet renormalization group analysis to study the stable background.
This program has been most successful for open string
tachyons living on unstable D-branes, where 
both open string field theory and boundary RG analysis support the intuitive
conjecture that the true vacuum corresponds to the closed string vacuum without
the branes \cite{Sen:2004nf}.
Closed string tachyons have been more challenging.
On the one hand closed string field theory is much
more complicated than open string field theory due to its 
non-polynomiality,\footnote{See \cite{Yang} 
for recent progress however.}
and on the other hand the worldsheet RG approach is generally hampered
by Zamolodchikov's $c$-theorem.
Conceptually, the question of closed string tachyons is expected
to be harder since closed string theories are theories of gravity, and
the dynamics of the tachyon is strongly correlated with the structure 
of spacetime. 
A simple analogy with open string tachyons suggests that the condensation
of closed string tachyons should lead to the decay of spacetime itself.
However we do not know what this means in the context of the original 
string theory.

The above discussion raises two questions. The first is whether
we can get a better handle on the fate of backgrounds with closed string
tachyons, and the second is whether there is a connection between tachyonic 
and semiclassical instabilities in closed string theory.

Recently some progress has been made on the first question in the context
of closed string tachyons which are localized on lower dimensional hypersurfaces
in space. 
There are a few examples of backgrounds with localized closed string tachyons
for which we have strong evidence regarding the nature of the tachyon
condensate \cite{Martinec:2002tz, Headrick:2004hz}. 
One example is the cone orbifold $\mathbb C/Z_N$ \cite{Adams:2001sv}.
Tachyons in the twisted sector of the orbifold are localized
at the tip of the cone, and it has been argued, based on a D-brane probe
analysis, that the condensation of these
tachyons leads to a reduction in the rank $N$, and thereby to a smoothing out
of the singularity. Further support for this picture
comes from worldsheet RG analysis \cite{Adams:2001sv,Vafa,Harvey:2001wm}, 
and closed string field theory \cite{Okawa:2004rh,Bergman:2004st}.
Another example consists of a SS circle which shrinks
locally in the transverse directions, leading to a localized winding
tachyon \cite{Adams:2005rb}. RG analysis suggests that the condensation
of this tachyon leads to a topology change in space, 
whereby the circle pinches and the transverse space splits at that point.
Localized winding tachyons may also play a role in resolving cosmological
singularities \cite{McGreevy:2005ci,Berkooz:2005ym}, black hole physics
\cite{Horowitz:2005vp,Ross:2005ms,Kutasov:2005rr},
and chronology protection \cite{Costa:2005ej}.
In both examples the effect of the tachyon is felt only locally, and the
bulk of spacetime remains intact. It would be nice to have an example
with a non-localized tachyon.\footnote{Attempts were made in
\cite{Strominger:2003fn,Schomerus:2003vv} for the bulk rolling tachyon,
\cite{Hirano:2005sg} for the bulk bouncing tachyon. See 
\cite{Suyama:2002ky} for more general observations.} 

The second question is interesting since, if true, it would demystify
somewhat the role of closed string tachyons.
Semiclassical instabilities provide a clear geometrical picture for
the decay of an unstable background, at least in the regime where
the low-energy approximation is valid.\footnote{This is also true in the
open string case \cite{Callan:1997kz, Hashimoto:2002xt}.}
De Alwis and Flournoy have conjectured
that when continued past this regime, the instability would  eventually
become tachyonic \cite{DeAlwis:2002kp}. As evidence for this they list several
examples of string backgrounds, such as the SS circle, which 
exhibit a semiclassical instability in one regime of the moduli space
($R>>1$), and a tachyonic instability in another regime ($R<1$).
In field theory we are used to such a transition.
Consider for example $\lambda\phi^4$ field theory with $\lambda<0$.
If $m^2>0$ the origin $\phi=0$ is a local minimum, which is unstable to decay
by tunneling out to $\phi=\pm m/\sqrt{-\lambda}$, and then rolling down.
If $m^2<0$ the origin is a local maximum, {\em i.e.} tachyonic,
and the field decays by rolling (Fig.~1).
In both cases the post-decay state is the same.
In closed string theory the picture is less clear because on the one
hand we don't  
have a good description of the tachyon condensate, and on the other hand 
the semiclassical regime is very far from the tachyonic
regime.

\begin{figure}[htbp]
\centerline{\epsfxsize=3in\epsfbox{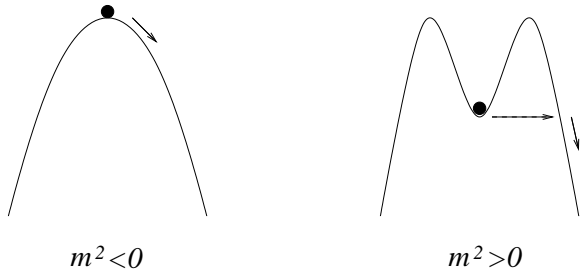}}
\medskip
\caption{Tachyonic and semiclassical instabilities in $\lambda\phi^4$ 
($\lambda<0$) field theory.}
\end{figure}

An approach which may shed some light on both questions is to focus
on the point in moduli space where the tachyon becomes massless.
At this point the tachyon condensate
would correspond to an on-shell state (if the corresponding vertex
operator is exactly marginal), and the true vacuum would be described
by an exact two-dimensional CFT. 
In fact the first hard evidence for Sen's conjecture about open string tachyons
came from this direction.
Sen considered an unstable D$p$-brane wrapped on a circle of radius $R$,
and a unit KK momentum ``kink'' mode of the open string tachyon \cite{Sen:1999mh}.
This mode becomes massless when $R=1/2$, and the corresponding operator
is exactly marginal. The condensate of this mode is described
by an exact CFT given by the boundary Sine-Gordon model, 
whose solution interpolates between Neumann and Dirichlet boundary conditions 
on the circle \cite{Polchinski:1994my,Callan:1994ub}, which in turn
implies that the tachyon kink corresponds to a D$(p-1)$-brane.
A simple closed string example of this is the winding tachyon in the
SS compactification of Type II string theory at $R=1$, or 
the bosonic string at $R=2$. The worldsheet field theory of the tachyon
condensate is given by the massless limit of the Sine-Gordon model.
However as far as we can tell, it is not known whether this theory
is an exact CFT. 

There exists a related theory which is an exact CFT, given by
the Sine-Liouville model. Motivated by this, and by the conjectured
duality of the Sine-Liouville model and the cigar $(SL(2,\mathbb C)/SU(2))/U(1)$
CFT \cite{FZZ,Kazakov:2000pm}, 
we set out to study the instability of the linear dilaton background
compactified on a circle (with SS boundary conditions in Type
II).\footnote{A related topic was discussed in \cite{Suyama:2002xk}.}
The Sine-Liouville model describes the massless limit, at $R=\sqrt{k}$, 
where $k$ is the level of $SL(2)$ current algebra in the dual CFT,
of a tachyon which condenses in an exponential profile, 
{\em i.e.} a semi-localized tachyon.
The dual cigar background clearly shows that the condensate corresponds
geometrically to a smooth truncation of the relevant two-dimensional part
of the space from an infinite cylinder to a semi-infinite 
cigar (Fig.~2). A similar decay was argued to occur in finite
temperature string theory on AdS below the Hagedorn temperature 
\cite{Barbon:2004dd,Kruczenski:2005pj}. An open string version of this was 
recently discussed in \cite{Kutasov:2005rr}.

\begin{figure}[htbp]
\centerline{\epsfxsize=4in\epsfbox{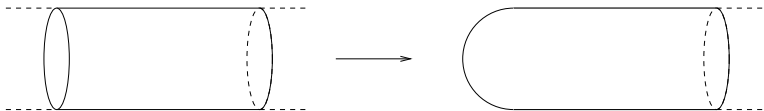}}
\medskip
\caption{Semi-localized tachyon condensation turns the cylinder into a cigar.}
\end{figure}

Interestingly, this example also gives an explicit connection between
a tachyonic instability and a semiclassical instability.
We will show that the cigar is a type of bubble spacetime which results from
a bounce, and that this can be generalized to larger radii
(but not smaller). For $R>\sqrt{k}$ the KK (or SS) linear dilaton background
is therefore unstable to decay semiclasically into a cigar-like
spacetime, and for $R<\sqrt{k}$ it is unstable due to a semi-localized tachyon.
The two instabilities are equivalent at $R=\sqrt{k}$.
Furthermore, since $k$ can be made large the supergravity approximation
remains valid all the way down to the transition point.

\medskip

The rest of the paper is organized as follows.
In section 2 we review the relevant features of the cigar,
the duality between the cigar and the Sine-Liouville model, and 
a three-dimensional deformation of the cigar.
In section 3 we put the Sine-Liouville model and the cigar in the
context of the instability of the KK (or SS) linear dilaton background.
We show that the cigar can be interpreted as a type of bubble of nothing
resulting from a gravitational bounce,
and generalize this to larger radii using a deformed cigar background. 
We then compute the classical decay rate due to bubble nucleation. 
In section 4 we show, using a generalized cigar-Sine-Liouville duality,
that the semiclassical decay of the KK linear dilaton via the gravitational
bounce can also be described as a field-theoretic tunneling process of the 
massive scalar field corresponding to the semi-localized tachyon
for $R>\sqrt{k}$. In section 5 we address the question of the negative mode
in the Euclidean backgrounds, which is relevant for the quantum-corrected
decay rate. Section 6 contains our conclusions.
We have also included an appendix containing a description of the spectrum
in the post-decay background.


\section{Review of some known results}

We begin by reviewing some known results which we will use in
the following sections. These include the two-dimensional cigar, the 
conjectured duality between the cigar and Sine-Liouville theory, and
a certain three-dimensional deformation of the cigar.

\subsection{The cigar}

The two-dimensional cigar is described by the metric and dilaton 
\cite{Witten:1991yr,Mandal:1991tz,Elitzur:1991cb}
\be
\label{cigar_metric}
 ds^2 &=& k\left[ dr^2 + \tanh^2r d\phi^2 \right] \nonumber\\
 \Phi-\Phi_0 &=& -\log\cosh r ,
\ee
where $\phi$ is a periodic coordinate with $\phi\sim\phi + 2\pi$.
In string theory the cigar corresponds to an exact conformal field theory
given by the coset $(SL(2,\mathbb C)/SU(2))/U(1)$, where the parameter $k$
in the metric corresponds to the level of the $SL(2)$ current algebra.
The central charge in the bosonic string case is given by
\be
 c = {3k\over k-2} - 1.
\label{c_cigar}
\ee
There exists also a supersymmetric version of the cigar CFT \cite{Giveon:1999px}
with a central charge $c=3(k+2)/k$, and an odd spin structure
at $r\rightarrow\infty$.

The basic set of local operators (ignoring the oscillators) 
is labeled by three quantum numbers $\{j,m,\bar{m}\}$,
where $j$ is associated with the dependence on $r$, and 
$m$ and $\bar{m}$ are related to the momentum and winding around the cigar as
\be
\label{m}
 m = {1\over 2}\left(n_\phi+kw_\phi\right) \;,\;
 \bar{m} = -{1\over 2}\left(n_\phi-kw_\phi\right) .
\ee
The conformal dimensions of these operators are given by
\be
 \Delta_{j,m,\bar{m}} = -{j(j+1)\over k-2} + {m^2\over k} \;,\;
 \bar{\Delta}_{j,m,\bar{m}} = -{j(j+1)\over k-2} + {\bar{m}^2\over k}.
\label{jmdimensions}
\ee
In the supersymmetric case $k-2$ is replaced by $k$ in the $j$ dependent part.

There are two kinds of states in this background.
First, there are bulk states which can propagate along the cigar. 
These are ordinary normalizable states which are the same as the states in   
the asymptotic linear dilaton background. In particular, they 
have a continuous spectrum with $j= -1/2 + is$, where $s\in\mathbb R$.
Second, there are states which are localized at 
the tip of the cigar with $j\in\mathbb R$ \cite{Dijkgraaf:1991ba}.
Unitarity places a bound on $j$ given by  $j<(k-2)/2$ 
($j<k/2$ in the supersymmetric case),
and since the formulas for the conformal dimensions (\ref{jmdimensions})
are invariant under the replacement $j\rightarrow -j-1$ one can restrict to
$j>-1/2$. These states are non-normalizable in general, however 
as argued in \cite{Giveon:1999px,Aharony:2004xn,Kutasov:2000jp} a subset of them given by 
\be
 -{1\over 2}<j<{k-3\over 2} \ , \ m=\pm(j+l) \ \Big(\mbox{or}\ \bar{m}=\pm(j+l)\Big) \ , \ 
l\in\mathbb Z_+
\ee
(and $k\rightarrow k+2$ in the supersymmetric case) are normalizable. 
Note that the upper bound on $j$ is tighter than the one imposed by unitarity.
The cigar CFT is closely related 
to the $SL(2,\mathbb R)$ WZW model, {\em i.e.} to $AdS_3$ \cite{Maldacena:2000hw}.
The bulk states correspond
to the continuous representations of $SL(2,\mathbb R)$ (the long strings),
and the normalizable localized states to the discrete representations 
(the short strings), including their images under spectral flow.

\subsection{FZZ duality}

It has been conjectured that the cigar CFT is equivalent to
the Sine-Liouville model \cite{FZZ,Kazakov:2000pm} 
(see also \cite{Fukuda:2001jd}) given by 
\be
 {\cal L} = {1\over 4\pi}\left[(\partial x_1)^2 + (\partial x_2)^2
    + Q\hat{R}x_1 + \lambda \, e^{bx_1}
\cos R x_2'\right],
\label{SL}
\ee
where $x_2'\equiv x_{2L}-x_{2R}$.
This is a linear dilaton CFT with
\be
 \Phi-\Phi_0 = -Qx_1,
\ee
and a Sine-Liouville interaction.
The central charge of this theory is $c=2+6Q^2$. Equating this with the
central charge of the cigar (\ref{c_cigar}) gives $Q^2=1/(k-2)$.
An analogous conjecture relates $N=2$ Liouville theory \cite{Kutasov:1990ua} to
the supersymmetric version 
of the cigar coset CFT
\cite{Hori:2001ax,Tong:2003ik,Eguchi:2004ik,Ahn:2002sx},
in which case $Q^2=1/k$.
In the Sine-Liouville model $x_2$ is a periodic coordinate with period $2\pi R$.
The radius is taken to be $R=\sqrt{k}$, in agreement with the asymptotic radius 
of the cigar (\ref{cigar_metric}). Requiring the Sine-Liouville potential to be
marginal then fixes $b=-1/Q$.
In the asymptotic weakly-coupled region both the cigar and Sine-Liouville model
look like a cylinder with a linear dilaton, and the coordinates are identified
(for large $k$) as $r\sim Qx_1$ and $\phi\sim x_2/\sqrt{k}$.

The local operators discussed above correspond in the Sine-Liouville
model to 
\be
 {\cal O}_{a,p_2} \sim e^{a x_1 + ip_{2L}x_{2L} + ip_{2R}x_{2R}}
\label{SLoperators}
\ee
with scaling dimensions
\be
 \Delta_{a,p_2} = -{1\over 4}a\left(a+2Q\right) + {1\over 4}p_{2L}^2\;,\;
 \bar{\Delta}_{a,p_2} = -{1\over 4}a\left(a+2Q\right)  + {1\over 4}p_{2R}^2 ,
\ee
where\footnote{This differs from \cite{Kazakov:2000pm} by the replacement 
$j\rightarrow -j-1$.} 
\be
 a = -2Q(j+1) \;,\;
 p_{2L} = {n_\phi + kw_\phi\over\sqrt{k}}  \;,\;
 p_{2R} = {n_\phi- kw_\phi\over\sqrt{k}} \ .
\ee
Bulk states therefore correspond to operators with $a = -Q + ip_1$,
exactly as in the free linear-dilaton theory, and the normalizable localized
states (localized near the Liouville wall) are given by
\be
-(Q+1/Q)<a<-Q \ , \ n_\phi + kw_\phi = \pm \left(-{a\over Q} + l - 1\right)
\ , \ l\in\mathbb Z_+ \ ,
\ee
or the same with $w_\phi\rightarrow -w_\phi$.
In particular the Sine-Liouville potential itself corresponds to the localized
state with $a=-1/Q$ (or equivalently $j=(k-4)/2$ in the bosonic string,
and $j=(k-2)/2$ in the superstring), $n_\phi=0$ and $w_\phi=1$.

\subsection{Deformed cigar}

A background which generalizes the cigar can be obtained by the $J^3\bar{J}^3$
deformation of $SL(2,\mathbb R)$ defined by the coset 
\cite{Horne:1991gn, Israel:2003ry}\footnote{The related $SU(2)$ case was discussed in
\cite{Hassan:1992gi,Giveon:1993ph, Yang:1988bi}.}
\be
\label{coset}
{SL(2,\mathbb{R})\times U(1)\over U(1)} \ .
\ee 
The corresponding metric, dilaton and $B$ field are given (in the string frame) by 
\be
\label{deformed_cigar}
ds^2&=&k\left[dr^2+
{\alpha\tanh^2r\over\alpha-\tanh^2r}
d\phi^2-{dt^2\over \alpha-\tanh^2r}\right] \nonumber\\
e^{2(\Phi-\Phi_0)}&=&{\alpha-1\over \cosh^2r(\alpha-\tanh^2r)}\\
B&=&k{\tanh^2r\over \alpha-\tanh^2r}d\phi\wedge dt \ ,\nonumber
\ee
where $\alpha$ is the deformation parameter.
The original $SL(2,\mathbb R)$ corresponds to $\alpha=1$.
In the limit $\alpha\rightarrow\infty$ the background reduces to 
$cigar\times\mathbb R$, and for $\alpha\rightarrow 0$ it becomes 
$trumpet\times\mathbb R$.
For $\alpha<0$ the geometry is Euclidean; it can be continued to a Lorentzian
spacetime by replacing $t\rightarrow it$, but then $B$ becomes imaginary,
so the background is not physical.

We will be interested in the geometries with $\alpha >1$,
which can be regarded as deformations of $cigar\times \mathbb R$.
The classical supergravity approximation is valid for $k\gg 1$ and 
$e^{2\Phi_0}\ll 1$. 
The asymptotic geometry is that of a cylinder 
with radius $R= \sqrt{k\alpha/(\alpha -1)}>\sqrt{k}$, 
a linear dilaton $\Phi(r) \sim  -r$ and 
$B \sim  {k\over \alpha -1}\, d\phi\wedge dt$.\footnote{The coefficient for the 
dilaton in (\ref{deformed_cigar}) was chosen so that the dilaton is independent 
of $\alpha$ in the limit $r\rightarrow\infty$.}
Note that the non-trivial asymptotic $B$ field will
affect the spectrum of winding states.
In our application we will consider a closely related background with 
a $B$-field whose $(\phi,t)$ component vanishes asymptotically,
so that the asymptotic background is the ordinary KK 
(or SS) linear dilaton.

\medskip

The background (\ref{deformed_cigar}) actually corresponds to an infinite 
cover of the coset (\ref{coset}).
For the single cover $t$ is periodic with 
$t\sim t + 2\pi$. We can define the $N$th cover by replacing 
$t\rightarrow Nt$, and the infinite cover corresponds to $N\rightarrow\infty$
with non-compact time \cite{Israel:2005ek}.

The spectrum of this background is most easily obtained by T-dualizing
along $t$ \cite{Israel:2005ek}. 
Since T-duality of a timelike coordinate is problematic we will
define it by T-dualizing 
the Euclidean geometry with $\alpha<0$, and analytically continuing to 
$\alpha>1$.\footnote{Alternatively, we can perform T-duality with respect to the
coordinate $it$.}
The T-dual background is given by\footnote{This
  differs by a  
sign from \cite{Israel:2005ek}.}
\be
\label{Tdual}
 ds^{\prime 2} &=& k\left[ dr^2 + \tanh^2r \left(d\phi - {dt'\over kN}\right)^2
 - \alpha\left({dt'\over kN}\right)^2\right] \nonumber\\
 \Phi'-\Phi'_0 &=& -\log\cosh r ,
\ee
and a trivial $B$ field.
Redefining $\widetilde{\phi}=\phi-t'/(kN)$ and $\widetilde{t}'=t'/(kN)$, 
we see that this is basically a $cigar\times S^1$, but
with the identifications
\be
 (\widetilde{\phi},\widetilde{t}')\sim 
\left(\widetilde{\phi} + 2\pi n - {2\pi n'\over kN},
\widetilde{t}'+{2\pi n'\over kN}\right)\quad
 \;n,n'\in\mathbb Z \ .
\ee
This is the orbifold $ (cigar\times U(1)_{-\alpha k})/\mathbb Z_{kN}$.
The spectrum is obtained by tensoring the cigar part of the spectrum 
$\{j,\widetilde{n}_\phi,\widetilde{w}_\phi\}$ 
with the $U(1)_{-\alpha k}$ part, labeled by momentum and winding numbers
$\{\widetilde{n}'_t,\widetilde{w}'_t\}$,
including twisted sectors labeled by $\gamma\in\mathbb Z_{kN}$,
and projecting onto the $\mathbb Z_{kN}$ invariant sector.
The invariant states must satisfy
\be
 \widetilde{n}'_t=\widetilde{n}_\phi + kNp \ , \quad p\in\mathbb Z \ .
\ee
The twist $\gamma$ corresponds to a fractional winding in $\widetilde{\phi}$,
as can be seen from the relations between the winding numbers 
in the $\widetilde{\phi}, \widetilde{t}'$ and $\phi, t'$ coordinates,
\be
\label{tilde_windings}
 \widetilde{w}_\phi = w_\phi - {1\over kN}w'_t \ , \ 
 \widetilde{w}'_t = {1\over kN}w'_t \ ,
\ee 
and from $\gamma = w'_t \ \mbox{mod} \ kN$.


\section{The decay of the KK linear dilaton}

\subsection{The background}

Our initial background is 
$\mathbb R^{1,1}\times S^1\times {\cal M}_{D-3}$
with a linear dilaton,
\be
 ds^2 &=& dx_1^2 + dx_2^2 - dt^2 + ds_{D-3}^2\nonumber\\
 \Phi - \Phi_0 &=& -Qx_1 \ ,
\label{LDbackground}
\ee 
where $x_2\sim x_2 + 2\pi R$.
We will consider both the bosonic string and Type II strings.
For Type II strings we impose anti-periodic boundary conditions on the
fermions, {\em i.e.} SS compactification, so that
spacetime supersymmetry is broken.
The three-dimensional part of the background given by $(x_1,x_2,t)$
corresponds to a free linear-dilaton CFT
\be
 {\cal L} = {1\over 4\pi}\left[(\partial x_1)^2 + (\partial x_2)^2
    - (\partial t)^2 + Q\hat{R}x_1 \right] \ ,
\label{LD}
\ee
with a central charge $c=3+6Q^2$ for the bosonic string,
and $c=9/2 + 6Q^2$ for Type II strings.
The $(D-3)$-dimensional space ${\cal M}_{D-3}$
corresponds to a CFT with $c=23-6Q^2$ for the bosonic string,
and to a SCFT with $c=21/2-6Q^2$ for Type II strings.
For example this can be a subcritical bosonic string theory with 
${\cal M}_{D-3}=\mathbb R^{D-3}$
and $Q=\sqrt{(26-D)/6}$, or critical bosonic or Type II strings
with ${\cal M}_{D-3}=\mathbb R^{D-6} \times S^3$ and 
$Q=1/\sqrt{N+2}$ ($Q=1/\sqrt{N}$ in Type II), where $N$ is the level of
the $SU(2)$ current algebra which defines the $S^3$.
In the Type II case this is the CHS model \cite{Callan:1991at}, which
corresponds to the near-horizon geometry of $N$ extremal NS5-branes
\cite{Aharony:1998ub}.

The spectrum of local operators in the oscillator ground state is given by
\be
 {\cal O} \sim e^{-ip_0t} {\cal O}_{a,p_2} {\cal O}_{D-3}\ ,
\label{LDoperators}
\ee
where ${\cal O}_{a,p_2}$ is the same as (\ref{SLoperators}),
and ${\cal O}_{D-3}$ is a ground state local operator in ${\cal M}_{D-3}$.
Bulk states, as before, correspond to operators with $a=-Q+ip_1$.
Operators with $a\in \mathbb R$ correspond to localized (or more
properly semi-localized) states, which in the absence of a potential
are always non-normalizable.

The background described by (\ref{LDbackground}) is not perturbatively well-defined
since the string coupling $g_s=\exp(\Phi)$ grows exponentially, and there is a 
strong coupling singularity (in the Einstein frame metric) at $x_1\rightarrow -\infty$.
The usual way in which this problem is evaded is by turning on a potential,
corresponding to a non-normalizable state,
that prevents strings from propagating into the strong coupling region.

\subsection{A semi-localized tachyon}

Consider the localized operator with
the Sine-Liouville potential at an arbitrary radius $R$ (and $b=-1/Q$),
\be
\label{sltachyon}
{\cal O}_{a,p_2} \sim e^{-x_1/Q}\cos Rx_2' \ .
\ee
This can be thought of as a particular mode of the winding number one
ground state of the string. In Type II string theory 
the SS periodicity conditions imply that 
the ground state is projected out at even winding number, but kept
at odd winding number. We can compute the mass of this state by requiring
that the total scaling dimension be 1 in the bosonic case and $1/2$ in the
Type II case. Using $Q^2=1/(k-2)$ for the bosonic string and $Q^2=1/k$
for the Type II string (since we will want to relate this to the
(deformed) cigar) we find
\be
 M^2 = R^2 - k \ .
\ee
The state associated with the Sine-Liouville potential is therefore tachyonic
for $R<\sqrt{k}$, indicating that the KK (or SS in the Type II case) 
linear dilaton background becomes perturbatively unstable below 
$R=\sqrt{k}$.\footnote{Of course the bosonic string is perturbatively 
unstable for all $R$ due to the winding number 0 tachyon. 
We are ignoring this mode here.}

The semi-localized tachyon (\ref{sltachyon}) describes a 
particular non-uniform mode
of the closed string tachyon, in which the tachyon condenses in the strong 
coupling regime $x_1<0$. At the critical radius $R=\sqrt{k}$ this mode
is exactly marginal, and the deformed CFT which describes the corresponding
condensate is given by the Sine-Liouville model (\ref{SL}) tensored with
$\mathbb R\times {\cal M}_{D-3}$.

The semi-localized tachyon background provides the potential which
prevents strings from propagating into the strong coupling region.
In this sense it effectively eliminates that part of the spacetime 
in which it condenses, which includes the singularity at $x_1\rightarrow -\infty$.
A more precise spacetime picture will emerge from the dual cigar CFT in the 
next section.

\subsection{Semi-classical instability at $R=\sqrt{k}$}

FZZ duality implies that the semi-localized tachyon condensate
at the critical radius $R=\sqrt{k}$ in the KK linear dilaton background 
is equivalent to the background 
given by $cigar\times\mathbb R\times{\cal M}_{D-3}$:
\be
\label{cigar_bubble}
 ds^2 &=& k\left( dr^2 + \tanh^2r d\phi^2 \right) -dt^2 + ds^2_{D-3}\nonumber\\
 \Phi-\Phi_0 &=& -\log\cosh r .
\ee
This provides a clear spacetime
interpretation of the semi-localized tachyon condensate, whereby
the half of spacetime containing the strong coupling region is excised.
But what is even more intriguing is that in the
dual picture the background is actually a type of ``bubble of nothing''.
To see this we introduce a new radial coordinate
\be
 \rho \equiv e^{-2(\Phi-\Phi_0)/(D-2)}=\left(\cosh r\right)^{2\over D-2}
\ee
and change to the Einstein frame
\be
 ds_E^2= \left({D-2\over 2}\right)^2 {k\over 1-\rho^{2-D}}\, d\rho^2
+ k \rho^2\left(1-\rho^{2-D}\right)d\phi^2 -\rho^2dt^2
+\rho^2ds_{D-3}^2\ .
\label{postdecayEin}
\ee
The size of the $\phi$-circle shrinks to zero as $\rho\rightarrow 1$, so
the region $0\leq\rho<1$ is excised from the space, forming a static 
``bubble of nothing''. 
The usual expanding bubble picture is obtained by changing to the coordinates
\be 
 u &=& {(D-2)\sqrt{k}\over 2}\rho\cosh\left[{2t\over (D-2)\sqrt{k}}\right] \nonumber\\
 v &=& {(D-2)\sqrt{k}\over 2}\rho\sinh\left[{2t\over (D-2)\sqrt{k}}\right] \ ,
\ee
in terms of which the $\rho,t$ part of the metric is asymptotically flat.
In these coordinates the surface of the bubble is at $u^2-v^2=(D-2)^2k/4$.
The bubble appears at time $v=0$, and expands with
time $v$ at a speed approaching the speed of light (Fig.~3).

\begin{figure}[htbp]
\centerline{\epsfxsize=2.5in\epsfbox{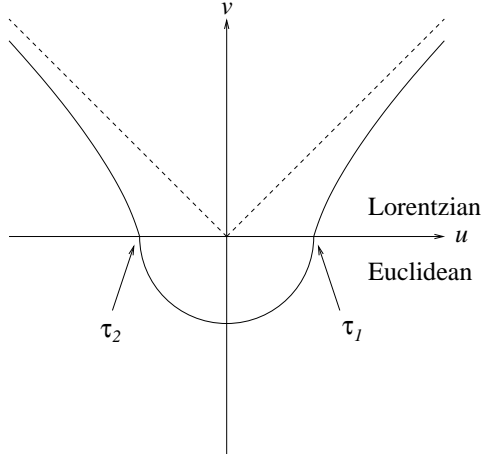}}
\medskip
\caption{Bubble nucleation and expansion.}
\end{figure}

This strongly suggests that the spacetime described by (\ref{cigar_bubble})
can be identified as the post-decay spacetime resulting from a bounce
in the KK linear dilaton background at $R=\sqrt{k}$. 
The bounce is described by the Euclidean geometry
\be
\label{cigar_bounce}
 ds^2 = k\left( dr^2 + \tanh^2r d\phi^2 \right) + d\tau^2 + ds^2_{D-3} \ ,
\ee
where $\tau = it$.
Single-valuedness in the $(u,v)$ plane requires a definite periodicity for $\tau$,
\be
 \tau \sim \tau + \pi  (D-2)\sqrt{k} \ .
\label{periodicity}
\ee
The bubble corresponds to the Lorentzian continuation of the Euclidean 
geometry at the turning points $\tau_1 =0$ and $\tau_2=\pi(D-2)\sqrt{k}/2$
(Fig.~3).
Note that unlike the ordinary KK bubble \cite{Witten:1981gj} 
the Lorentzian geometry here actually consists of two identical disconnected
components. We should therefore think of the original background as two copies
of the KK linear dilaton, each of which decays into one of these components
(Fig.~4).

\begin{figure}[htbp]
\centerline{\epsfxsize=2.5in\epsfbox{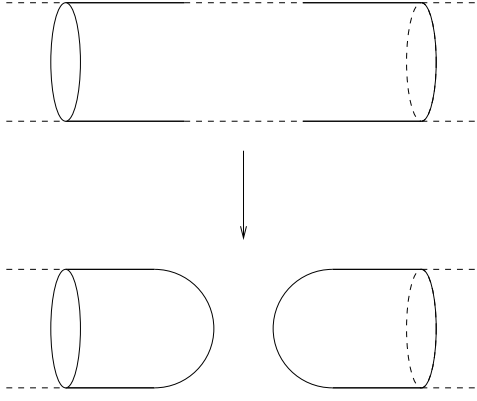}}
\medskip
\caption{Two copies of the cylinder decay into two cigars.}
\end{figure}

\subsection{Generalization to $R>\sqrt{k}$}

For $R>\sqrt{k}$ the Sine-Liouville operator (\ref{sltachyon}) is irrelevant, 
corresponding to a massive state. In this case the KK linear dilaton background is
perturbatively stable. On the other hand we can deform the three-dimensional
$cigar\times \mathbb R$ part of the spacetime (\ref{cigar_bubble}) to a spacetime 
(\ref{deformed_cigar}) with $R>\sqrt{k}$, which suggests a generalization
of the cigar bounce instability.
We are actually interested in a slight variation of the deformed
background given by\footnote{We are suppressing the components of the $B$ field 
on the $S^3$.}
\be
\label{deformed_bubble}
ds^2&=&k\left[dr^2+{\alpha\tanh^2r\over\alpha-\tanh^2r}\, d\phi^2
 -{N^2\over \alpha-\tanh^2r}\, dt^2 \right] + ds^2_{D-3}\nonumber\\
e^{2(\Phi-\Phi_0)}&=&{\alpha-1\over \cosh^2r(\alpha-\tanh^2r)}\\
B&=&Nk\left[{\tanh^2r\over \alpha-\tanh^2r} - {1\over \alpha-1}\right]\,
d\phi\wedge dt \ , \nonumber
\ee
where $\phi\sim \phi + 2\pi$ and $t\sim t + 2\pi$.
As we explained in the previous section, this corresponds to the $N$-fold 
cover of the coset $(SL(2,\mathbb R)\times U(1))/U(1)$.
We will eventually take $N\rightarrow\infty$ to get non-compact time.
We have also added a constant term to the $(\phi,t)$ component of the $B$ field
so that it vanishes at $r\rightarrow\infty$.
This background asymptotes to 
$\mathbb R \times S^1 \times S^1_t\times {\cal M}_{D-3}$
with a linear dilaton, $R^2 = k\alpha/(\alpha-1)$ and
$R_t^2 = kN^2/(\alpha-1)$. 

The Einstein metric is given by
\be
 ds_E^2 &=& \left({D-2\over 2}\right)^2 
{(\alpha-1)^2 k\over (\alpha -1-\alpha\rho^{2-D})(\alpha-1-\rho^{2-D})}\, d\rho^2
+ {\alpha k\over (\alpha -1)^2} \,
   \rho^2\left(\alpha-1-\alpha\rho^{2-D}\right)d\phi^2 \nonumber\\
  && \mbox{} - {kN^2\over (\alpha-1)^2}\, \rho^2(\alpha-1-\rho^{2-D}) dt^2
+\rho^2ds_{D-3}^2\ ,
\label{postdecayEin2}
\ee
where
\be
 \rho \equiv e^{-2(\Phi-\Phi_0)/(D-2)} = 
  \left[{(\alpha-1)\over \cosh^2r(\alpha - \tanh^2r)}\right]^{-2/(D-2)}\ ,
\ee
so the surface of the bubble is now at $\rho=[\alpha/(\alpha-1)]^{1/(D-2)}$.
We can again define 2d asymptotically flat coordinates
\be 
 u &=& {(D-2)\sqrt{k}\over 2}\rho\cosh\left[{2Nt\over 
(D-2)\sqrt{\alpha-1}}\right] \nonumber\\
 v &=& {(D-2)\sqrt{k}\over 2}\rho\sinh\left[{2Nt\over (D-2)\sqrt{\alpha-1}}\right] \ ,
\ee
such that the surface of the bubble expands in time $v$.

The bounce is given by the Euclidean continuation of
(\ref{deformed_bubble}) with $\tau = iNt$,
\be
\label{deformed_bounce}
ds^2&=&k\left[dr^2+{\alpha\tanh^2r\over\alpha-\tanh^2r}\, d\phi^2
 +{1\over \alpha-\tanh^2r}\, d\tau^2 \right] + ds^2_{D-3}\nonumber\\
B&=&-i k\left[{\tanh^2r\over \alpha-\tanh^2r} - {1\over \alpha-1}\right]\,
d\phi\wedge d\tau \ , 
\ee
and the same dilaton.
Single-valuedness in the $(u,v)$ coordinates now requires a periodicity
\be
\label{periodicity2}
 \tau \sim \tau + 2\pi \beta \ , \quad \beta\equiv {(D-2)\sqrt{\alpha-1}\over 2} \ .
\ee
Note that the asymptotic radius in the $\tau$ direction 
is given by $R_\tau = (D-2)\sqrt{k}/2$, independent of $\alpha$.
Here the continuation makes $B$
complex (the $S^3$ part is real, and the $\phi,t$ part is imaginary),
but this is acceptable in a Euclidean configuration 
\cite{Brown:1990fk, Hawking:1995ap, Booth:1998pb}.
We therefore claim that (\ref{deformed_bubble}) corresponds to the
post-decay background of the KK linear dilaton for $R>\sqrt{k}$.

Note that the Euclidean geometry (\ref{deformed_bubble}) with $\alpha<0$ is also 
asymptotic to a cylinder, this time with radius $R=\sqrt{k|\alpha|/(|\alpha|+1)}$,
which is smaller than $\sqrt{k}$.
Upon analytically continuing to Lorentzian signature by replacing $t\rightarrow it$,
this seems to give a deformation of the cigar to $R<\sqrt{k}$,
and therefore a bounce instability in the regime where the semi-localized 
state (\ref{sltachyon}) is tachyonic. Now, however,
the Lorentzian background has a complex $B$ field,
so this is not a physical background. 
This is consistent with our expectation that for $R<\sqrt{k}$ 
there should not be a bounce instability.

\subsection{Classical decay rate}

The classical rate of decay is estimated as $\exp(-I)$, where $I$ is the 
Euclidean action of the bounce \cite{Coleman:1977py}. 
To be specific we will consider the cases of the bosonic and Type II strings 
on $\mathbb R^{1,1}\times S^1\times S^3\times T^{D-6}$,
where the $(D-6)$-dimensional part is compactified on a torus
with volume $V_{D-6}$ in order to have a finite action.
The radius of the $S^3$ is given by the level $N$ of the $SU(2)$,
which for large $k$ is $N\sim k$.
The Euclidean metric which defines the bounce is given in this case by
\be
 ds^2 = k\left[dr^2+{\alpha\tanh^2r\over\alpha-\tanh^2r}\, d\phi^2
 + {1\over \alpha-\tanh^2r}\, d\tau^2 + d\Omega_3^2\right] 
 + \sum_{i=D-6}^{D-1} dy_i^2 \ ,
\ee
where $\phi\sim\phi+2\pi$, $\tau\sim\tau + 2\pi\beta$, and
$d\Omega_3$ is the volume form of $S^3$. 
The total
Euclidean three form field strength is given by
\be
H_3=2k\,d\Omega_3\ + i\,\frac{2k\alpha\tanh r\,
\mbox{sech}^2r}{(\alpha-\tanh^2 r)^2}\,
dr\wedge d\phi\wedge d\tau \ .
\ee
Since there are no background RR fields in the Type II case the
bulk part of the action is given by
\be
I_{bulk}=-{1\over 2\kappa^2}\int d^{D}x\sqrt{G}e^{-2\Phi}
\left[R+4|\partial\Phi|^2 -{1\over 2\cdot 3!}|H_3|^2\right]
\ee
for both the bosonic and Type II strings.

In evaluating the Euclidean action for the bounce
two kinds of boundary terms need to be included. The first is the Gibbons
Hawking term \cite{Gibbons:1976ue, York:1972sj}, 
which cancels the second derivative terms in the Einstein action,
\be
I_{GH}=-{1\over\kappa^2}
\int d^{D-1}x\sqrt{h_E}(K_E-K_{E}^{(0)})\ .
\ee
The subscript $E$ denotes that these quantities are defined
with respect to the Einstein frame.
$h_E$ is the induced metric on the boundary at $r\to\infty$, 
$K_E$ is the extrinsic curvature of the boundary 
\be
 K_E = \sqrt{G^{rr}_E}\,{\partial_r\sqrt{h_E}\over\sqrt{h_E}}\ ,
\ee
and $K_E^{(0)}$ is the extrinsic curvature of the
boundary in the background metric (the metric for the asymptotic geometry).
The second boundary term comes from the Euclidean action of the background metric
$I_{bg}$, which serves as a regularization term
(see for example \cite{Ng:2001ut}).\footnote{That this is a boundary term
follows from the equations of motion. For the same reason $I_{bulk}$
also gets a contribution only from the boundary.}
The regularized Euclidean action is thus
\be
I = I_{bulk}-I_{bg}+I_{GH} \ .
\ee
The computation is simplified somewhat using the equations of motion,
which give
\be
 R+4|\partial\Phi|^2 = {5\over 12} |H_3|^2 \ .
\ee
The different terms are given by
\be
I_{bulk}&=& -{8\pi^4(D-2)V_{D-6}\,k^2\over\kappa^2}\sqrt{\alpha\over \alpha -1}
\left[\cosh^2r - {\alpha\over\alpha-\tanh^2r}\right]_0^\infty \\[5pt]
I_{bg} &=& -{2\pi^4(D-2)V_{D-6}\,k^2\over\kappa^2}\sqrt{\alpha\over \alpha -1}
\Big[e^{2r}\Big]_{-\infty}^{\infty}\\[5pt]
I_{GH} &=& {4\pi^4 (2D-3) V_{D-6}\, k^2 \over (D-2)\kappa^2}
\sqrt{\alpha\over\alpha -1}\, {\alpha+1\over\alpha-1}\ .
\ee
Putting it all together we find that the rate of decay is given by
\be
\label{decay_rate}
 \Gamma \sim \exp\left[-{4\pi^4 (D-1)^2 V_{D-6} k^2\over (D-2)\kappa^2}  
\sqrt{\alpha\over\alpha -1}\, {\alpha+1\over\alpha-1}\right] 
\sim \exp\left[-cR(2R^2-k)\right] .
\ee

We would like to make two comments about this result.
First, the decay rate vanishes in the limit
$\alpha\rightarrow 1$, which
corresponds to an infinite radius. This is consistent with the fact
that the background is stable in this limit; for Type II strings
the background becomes supersymmetric, and therefore absolutely stable,
and for the bosonic string it is stable up to the usual winding number
0 tachyon.\footnote{Note that the two limits
$r\to\infty$ and $\alpha\to 1$ do not commute. 
The limit $\alpha\to 1$ of (\ref{deformed_bubble}) with $r$ finite is 
$AdS_3\times{\cal M}_{D-3}$. 
Here we first take $r\to\infty$ and then $\alpha\to 1$, which corresponds
to our interpretation of
the $\alpha\to 1$ limit as the post-decay geometry for 
$\mathbb R^{1,1}\times S^1\times{\cal M}_{D-3}$ with $R\rightarrow\infty$.}

Second, in the limit $\alpha\rightarrow\infty$, corresponding to 
the critical radius $R=\sqrt{k}$, the decay rate is still exponentially small.
This is somewhat surprising since this is the point where the semi-localized
tachyon appears, and we would expect an ${\cal O}(1)$ decay rate.
The same thing happens in the decay of the ordinary KK vacuum \cite{Witten:1981gj},
namely the decay rate remains exponentially small at the critical radius.
But there the semi-classical description breaks down at the critical radius,
which is $R\sim{\cal O}(1)$, whereas in our case the semi-classical picture
continues to hold as long as $k\gg 1$.
We will come back to this point in the next section.

\section{Tachyon bounce}

The picture that emerges is that 
the KK linear dilaton background is unstable to
the decay of half of spacetime for all $R$.
For $R<\sqrt{k}$ the instability is described by a semi-localized
tachyonic mode, and for $R>\sqrt{k}$ it corresponds to a gravitational bounce.
The two descriptions meet at $R=\sqrt{k}$, where they 
are related by FZZ duality. 
The situation is analogous to the $\lambda\phi^4$ 
field theory model with $\lambda < 0$.
The $\phi=0$ state is unstable to decay via rolling for $m^2<0$, 
and via tunneling for $m^2>0$.

In the related setting of finite temperature string theory,
Barbon and Rabinovici have argued that at temperatures below
the Hagedorn temperature, the background would decay into
a bubble of nothing by a tunneling process of the massive thermal tachyon
\cite{Barbon:2004dd}.
This can be put on firm footing in the context of the KK linear dilaton.

Consider first the critical radius $R=\sqrt{k}$.
In this case the dual descriptions of the post-decay background
are the cigar-bubble (\ref{cigar_bubble}) on one side, and the static tachyon
configuration
\be
 T \sim e^{-x_1/Q} \cos(\sqrt{k}x_2')
\ee
on the other.
For large $x_1$ we can express this configuration in terms of the 
asymptotically flat coordinates $(u,v)$ as
\be
\label{tachyon_evolution}
 T \sim (u^2 - v^2)^{-(D-2)\over 2Q^2}\cos(\sqrt{k}x_2')\ ,
\ee
which is a time-dependent configuration.
The bounce corresponds to $v\rightarrow iv_E$
\be
 T_E \sim (u^2 + v_E^2)^{-(D-2)\over 2Q^2}\cos(\sqrt{k}x_2')\ .
\ee
This is a tunneling process. At $v_E\rightarrow -\infty$ the field $T_E$ vanishes,
and at the turning point $v_E=0$ it takes its maximal value.
This is where the Lorentzian evolution (\ref{tachyon_evolution}) takes over,
and the field $T$ proceeds to increase with time $v$.
In this case the field is actually massless, so apparently this describes 
some sort of barrier-less tunneling (see \cite{Lee:1985uv} for field theory examples
of barrier-less tunneling). This is also consistent with the fact that the classical rate
of decay (\ref{decay_rate}) was exponentially suppressed even at the critical radius.

For $R>\sqrt{k}$ the post-decay background is given by (\ref{deformed_bubble}).
Since in the T-dual picture this can again be described as
a cigar, modulo a certain identification (see Appendix), we can use FZZ duality again
to get a dual tachyon configuration:
\be
 T \sim e^{-x_1/Q} \cos\left(\sqrt{k}\,\widetilde{x}_2'\right) \ ,
\ee
where the coordinate $\widetilde{x}_2$ is related asymptotically to the
cigar coordinate $\widetilde{\phi}$ as $\widetilde{x}_2\sim\sqrt{k}\, \widetilde{\phi}$.
Using (\ref{tilde}) we can express this in terms of the asymptotic flat
KK linear dilaton (and time) coordinates as
\be
\label{defSLpot}
 T \sim e^{-x_1/Q} \cos\left(\sqrt{\alpha k\over \alpha-1}x_2' -
 \sqrt{k\over\alpha-1} x_0\right) \ ,
\ee
where $x_2\sim R\phi = \sqrt{k\alpha/(\alpha-1)}\, \phi$, 
$x_0\sim R_t t = \sqrt{kN^2/(\alpha-1)}\, t$.
In terms of the $(u,v)$ coordinates this becomes
\be
\label{tachyon_evolution_2}
 T \sim (u^2 - v^2)^{-(D-2)\over 2Q^2} 
  \cos\left(\sqrt{\alpha k\over \alpha-1} x_2' - {D-2\over 2}{k\over\sqrt{\alpha-1}}
 \tanh^{-1}{v\over u}\right)\ .
\ee
The Euclidean continuation is complex
\be 
\label{tachyon_bounce_2}
T_E \sim (u^2 + v_E^2)^{-(D-2)\over 2Q^2} 
  \cos\left(Rx_2' - i {D-2\over 2}{k\over\sqrt{\alpha-1}}  \tan^{-1}{v_E\over u}\right)\ .
\ee
This can also be seen in the original deformed background (\ref{deformed_bubble}),
where the $B$ field becomes complex after Wick-rotation, or in the T-dual
background (\ref{Tdual2}), where the metric becomes complex.
The bounce in this case is not time-reversal symmetric, since the imaginary part
of (\ref{tachyon_bounce_2}) is odd under time reversal.
This is similar to a system with friction. The friction term in the action 
breaks time-reversal symmetry, and in the Euclidean theory it becomes
imaginary, rendering the solution complex.
The post-tunneling evolution is given by analytically continuing the Euclidean
solution at $v_E=0$ to the Lorentzian solution (\ref{tachyon_evolution_2})
at $v=0$. An unusual property in this case is that the evolution begins with a 
non-zero velocity, which is associated with the imaginary part of the Euclidean
solution.\footnote{A similar thing happens in the instanton
process representing the Schwinger pair production in the presence
of constant electric and magnetic fields. 
The instanton solution in that case is also complex, 
and the electron-positron pair are created with a non-zero velocity
due to the magnetic field.}

\section{Where is the negative mode?}

We have seen that the background (\ref{deformed_bubble}) is asymptotic
to the KK linear dilaton with $R\geq\sqrt{k}$, 
and we have computed the action of its Euclidean
continuation (\ref{deformed_bounce}). 
However a crucial requirement for the Euclidean background 
to correspond to a bounce is the existence of a negative mode in its
spectrum. This is because the rate
of decay including the one-loop contribution is \cite{Callan:1977pt} 
\be
\label{oneloop}
 \Gamma \sim \mbox{Im}\left[ 
{\mbox{det}^{1/2} \Delta_{bounce}\over\mbox{det}^{1/2} \Delta_{0}}\right]\,
e^{-I}\ ,
\ee
where $\Delta_{bounce}$ and $\Delta_0$ are the quadratic forms of the fluctuations
in the bounce and in the Euclidean continuation of the pre-decay background, respectively.
So we need to determine if the Euclidean background (\ref{deformed_bounce}) possesses
a negative mode, which the Euclidean continuation of the  KK linear dilaton background
does not. A negative mode corresponds to an operator with (holomorphic) scaling dimension 
less than 1 in the bosonic case, and less than $1/2$ in the Type II case.

The spectrum of the Euclidean background can be determined formally 
as in the Lorentzian case (see Appendix), by relating it via T-duality 
along $\tau$ to a cigar.
The resulting scaling dimensions can be read off from (\ref{spectrum})
by replacing 
\be
\label{replacements}
 n'_t\rightarrow -i\beta n'_\tau/N \ , \ w'_t\rightarrow iw'_\tau N/\beta \ ,
\ee
which gives 
\be
\label{Espectrum}
 \Delta + \bar{\Delta} &=& -{2j(j+1)\over k-2}
  + {\alpha-1\over 2\alpha k}n_\phi^2 
 + {\alpha k \over 2(\alpha-1)}w_\phi^2
  + {k(D-2)^2\over 8} n^{\prime 2}_\tau
  + {2\over k(D-2)^2} w^{\prime 2}_\tau \ . \nonumber \\
\ee
For the bulk states, {\em i.e.} the states with
$j=-1/2 +is$, this is the correct thing to do.
In particular this gives normalizable states with real scaling dimensions. 
This is exactly the same spectrum of bulk states as in the Euclidean continuation
of the pre-decay KK linear dilaton background, and so will not contribute in 
(\ref{oneloop}). The required negative mode must therefore come from a localized
state. The problem is that the replacements (\ref{replacements}) generally give complex
scaling dimensions for the localized states, as can be inferred from the relations
(\ref{localized_states}), (\ref{momenta}) and 
(\ref{windings}).\footnote{A similar thing happens to the discrete series 
in the Euclidean continuation of $AdS_3$ \cite{Maldacena:2001km}.
The $SL(2,\mathbb R)$ quantum numbers are related as 
$m=\pm(j+l)$ and $\bar{m}=\pm(j+\bar{l})$, where $l,\bar{l}\in\mathbb Z_+$.
On the other hand
$m$ and $\bar{m}$ are related to the energy and angular momentum in $AdS_3$ as
$m = (E+P_\theta)/2$ and $\bar{m} = (E-P_\theta)/2$.
The replacement $E\rightarrow iE$ therefore makes $j$, and with it
the scaling dimension, complex for these states.}

We have not resolved this problem, and it is not completely clear to us whether and how 
the negative mode appears in the spectrum of localized states.
However let us outline a number of proposals which may lead to a resolution.
\begin{enumerate}
\item States with trivial momentum and winding in $\tau'$
($n'_\tau=w'_\tau=0$) have real scaling dimensions. 
For $\alpha\gg 1$, and $j$ taking its maximal allowed 
value\footnote{More precisely, one needs $\alpha>k\gg 1$ for the
  value of $j$ not to exceed the upper bound. For smaller $\alpha$,
  one must choose larger $l$ in (\ref{localized_states}). For a given
  $l$, $\alpha$ has to be greater than $k/(2l-1)$.} 
\be
 j \sim \left\{
\begin{array}{ll}
 {k-4\over 2} + {k\over 2\alpha} & \ \mbox{for the bosonic string} \\[5pt]
 {k-2\over 2} + {k\over 2\alpha} & \ \mbox{for the Type II string} 
\end{array}
\right.
\ee
the ground state with $n_{\phi}=0$ and $w_{\phi}=1$ has the scaling
dimension
\be
\Delta\sim \left\{
\begin{array}{ll}
  1-{k-2\over 4\alpha} & \ \mbox{for the bosonic string}\\[5pt]
 \half -{k\over 4\alpha} & \ \mbox{for the Type II string}\ .
\end{array}
\right.
\ee
This corresponds to a negative mode.
In the cigar limit $\alpha\to\infty$ it becomes a zero mode, so the
one-loop decay rate vanishes. 
We expect that higher loop corrections render the decay rate finite
in this limit. 

\item For $\alpha\sim{\cal O}(1)$ the above mode is positive, since the
winding contribution to the scaling dimension grows indefinitely as 
$\alpha\rightarrow 1$, whereas the $j$ dependent part is ${\cal O}(k)$.
There can however be negative modes with $w_\phi=0$ at the first 
excited level (in the bosonic string there are additional negative
modes at the ground state).
For example the state with $n_{\phi}=k$ and $j=(k-2)/2$ in the Type II case
has a scaling dimension
\be
 \Delta = 1 - {k\over 4\alpha}\ ,
\ee
which corresponds to a negative mode if $\alpha<k/2$.
It is tempting to conjecture that this state supplies the required negative mode
for small $\alpha$, and the winding state above supplies the negative
mode for large $\alpha$.

\item An alternative possibility, 
following a similar proposal in \cite{Maldacena:2001km}, is to drop 
the $i$ in the replacements (\ref{replacements}) for the localized states.
This gives non-normalizable states with real scaling dimensions
in the Euclidean theory.
It isn't difficult to see that one can get arbitrarily negative 
scaling dimensions with large $n'_\tau$ or $w'_\tau$. 
The problem with this proposal in our case is that the 
corresponding vertex operators 
$\sim\exp[n'_\tau(\tau'_L+\tau'_R)+w'_\tau(\tau'_L-\tau'_R)]$
do not seem to be consistent with the periodicity of the
compact Euclidean time $\tau$ (\ref{periodicity2}). 
\end{enumerate}


\section{Conclusions}

The Kaluza-Klein compactification of the bosonic string,
and similarly the Scherk-Schwarz compactification of Type II strings, 
in a linear dilaton background, exhibits a semi-localized instability,
described by a semi-localized winding tachyon when $R\leq\sqrt{k}$, 
and by a gravitational bounce when $R\geq\sqrt{k}$.
At $R=\sqrt{k}$ the tachyon condensate is described by the Sine-Liouville
model, and the bounce corresponds to the cigar CFT.
The conjectured equivalence of the two models provides the first
explicit connection between a tachyonic instability and a semiclassical
instability, and implies that the tachyon condensate (at the critical radius)
corresponds to a smooth semi-infinite truncation of space.
The geometrical picture extends to $R>\sqrt{k}$, where the bounce
corresponds to a three-dimensional deformation of the cigar.

In this paper we have not discussed much the regime $R<\sqrt{k}$, where the decay
proceeds via a condensation of the semi-localized tachyon.
The dynamical evolution of the condensation will be described by a
rolling tachyon. The tachyon profile can be guessed 
from the $R>\sqrt{k}$ case (\ref{defSLpot}). 
We propose that it takes the form  
$T \sim \lambda e^{-x_1/Q} \cos(\sqrt{\alpha k/(\alpha-1)}x_2' 
-\sqrt{k/(\alpha-1)}x_0)+\mbox{c.c.}$
with $\alpha<0$. 
However, this represents an s-brane-like \cite{Gutperle:2002ai} process
rather than a rolling process.
To obtain a rolling process, one shifts 
$x_0\to x_0+\log\lambda/\sqrt{k/(|\alpha|+1)}$ 
and then takes the limit $\lambda\to 0$.
We believe that $T$ is not only marginal but an exactly marginal
operator, and the corresponding CFT is tractable by properly extending
the results of \cite{Baseilhac:1998eq}.
It is not clear if FZZ duality extends to this case.
However, it is tempting to speculate that the rolling tachyon does
have a spacetime dual description where the spacetime is time
dependent and consists of two copies of a cigar whose tips recede as
time goes by. The entire spacetime eventually disappears in the
infinite future, which is where the tachyon condensate becomes infinite.   


\section*{Acknowledgments}
We would like to thank Amit Giveon, Carlos Herdeiro, Dan Israel
and Harald Nieder for useful discussions.
This work is supported in part by the
Israel Science Foundation under grant no.~101/01-1.

\appendix
\section{The post-decay spectrum}

The spectrum of our deformed cigar background 
(\ref{deformed_bubble}) 
can be determined along the lines of \cite{Israel:2005ek}, by T-dualizing
along $t$. The T-dual background is given by
\be
\label{Tdual2}
 ds^{\prime 2} &=& k\left[ dr^2 + \tanh^2r 
  \left({\alpha d\phi\over\alpha-1} - {dt'\over kN}\right)^2
  - \alpha\left({dt'\over kN} - {d\phi\over\alpha-1}\right)^2\right] 
+ ds_{D-3}^2 \nonumber\\
 e^{2(\Phi'-\Phi'_0)}&=&{1-\alpha\over kN^2\cosh^2r} \ ,
\ee
and a trivial $B$ field. 
This differs a bit from the T-dual background in section 2.3 (\ref{Tdual})
due to the difference in the original $B$ field, but
it can again be related to a $cigar\times U(1)$ CFT by defining
\be
\label{tilde}
 \widetilde{\phi} \equiv {\alpha\phi\over\alpha-1} - {t'\over kN} \  , \ 
 \widetilde{t}' \equiv {t'\over kN} - {\phi\over\alpha-1} \ ,
\ee
which now have the identifications
\be
\label{identifications}
 (\widetilde{\phi},\widetilde{t}')\sim 
  \left(\widetilde{\phi} + {2\pi\alpha n\over \alpha-1} - {2\pi n'\over kN}  \, ,\,
            \widetilde{t}' + {2\pi n'\over kN} - {2\pi n\over \alpha-1} 
        \right) , \quad n,n'\in\mathbb Z \ .
\ee
With these identifications the background does not correspond to a simple orbifold
of the $cigar\times U(1)$ CFT, but we can still determine the spectrum
by imposing invariance of the $cigar\times U(1)$ spectrum.
By FZZ duality we know that the vertex operators are asymptotically given by
\be
 e^{-2(j+1)r}\, e^{i(\widetilde{n}_\phi + k\widetilde{w}_\phi)\widetilde{\phi}_L 
  + i(\widetilde{n}_\phi - k\widetilde{w}_\phi)\widetilde{\phi}_R}\, 
  e^{-i(\widetilde{n}'_t+k\alpha \widetilde{w}'_t)\widetilde{t}'_L 
- i(\widetilde{n}'_t-k\alpha \widetilde{w}'_t)\widetilde{t}'_R} \ ,
\ee
where $\widetilde{n}_\phi$ and $\widetilde{w}_\phi$ are the KK momentum and winding
number, respectively, in $\widetilde{\phi}$, and $\widetilde{n}'_t$ 
and $\widetilde{w}'_t$ are the KK momentum and winding number in $\widetilde{t}'$.
Invariance under the identifications in (\ref{identifications}) requires
\be
\label{momenta}
\widetilde{n}_\phi = n_\phi - {kN n'_t\over \alpha-1}\ , \ 
\widetilde{n}'_t = - n_\phi +{\alpha kN n'_t\over \alpha-1}\ ,
\ee
where $n_\phi,n'_t\in\mathbb Z$.
As we shall soon see $n_\phi$ and $n'_t$ will be identified as the KK momenta
in $\phi$ and $t'$, respectively. From (\ref{tilde}) we see that the winding numbers
are related as
\be
\label{windings}
 \widetilde{w}_\phi = {\alpha\over\alpha-1}w_\phi - {1\over kN}w'_t \ , \
 \widetilde{w}'_t = {1\over kN}w'_t - {1\over \alpha-1}w_\phi \ .
\ee 
Note in particular that in the limit $\alpha\rightarrow\infty$ this reduces to
the $\mathbb Z_{kN}$ orbifold, where the twist is identified with 
$w_t' \ \mbox{mod} \ kN$ (\ref{tilde_windings}). So for $w'_t, w_\phi \in\mathbb Z$ we believe this
describes the complete spectrum of the theory for any $\alpha > 1$ (including
the twisted sectors).
The total scaling dimension in the $cigar\times U(1)$ is given by (in the bosonic string)
\be
\label{spectrum}
 \Delta + \bar{\Delta} = -{2j(j+1)\over k-2}
  + {\alpha-1\over 2\alpha k}n_\phi^2 
  + {\alpha k \over 2(\alpha-1)}w_\phi^2
  - {kN^2\over 2(\alpha -1)}n_t^{\prime 2}
  - {\alpha-1\over 2kN^2}w^{\prime 2}_t 
\ee
and 
\be
 \Delta - \bar{\Delta} = n_\phi w_\phi - n'_t w'_t \ .
\ee
We can therefore identify $n_\phi$ as the KK momentum in $\phi$, and $n'_t$
as the KK momentum in $t'$, or equivalently as the winding number in the original $t$.
So the compact part of the spectrum is just what one would expect
for the asymptotic $S^1\times S^1_t$ geometry with $R=\sqrt{k\alpha/(\alpha-1)}$ and
$R_t=\sqrt{kN^2/(\alpha -1)}$.

The spectrum includes bulk and localized states. In terms of the cigar quantum numbers
the bulk states correspond to $j=-1/2+is$, and arbitrary 
$\widetilde{n}_\phi,\widetilde{w}_\phi$, and the localized states correspond
to real $j$ in the range $-1/2<j<(k-3)/2$ ($(k-1)/2$ in Type II), and
$\widetilde{n}_\phi,\widetilde{w}_\phi$ are constrained by
\be
\label{localized_states}
 \widetilde{n}_\phi + k\widetilde{w}_\phi = \pm 2(j+l) \quad \mbox{or} \quad
 \widetilde{n}_\phi - k\widetilde{w}_\phi = \pm 2(j+l) \ , \ 
 l\in\mathbb Z_+ \ .
\ee

\end{document}